\documentclass[conference]{IEEEtran}
\IEEEoverridecommandlockouts
\usepackage{cite}
\usepackage{amssymb,amsfonts}
\usepackage{algorithmic}
\usepackage{graphicx}
\usepackage{textcomp}
\usepackage{xcolor}
\usepackage{siunitx}
\usepackage{units}

\usepackage[T1]{fontenc}
\usepackage[utf8]{inputenc}
\usepackage{pgfplots}
\usepackage{grffile}
\pgfplotsset{compat=newest}
\usetikzlibrary{plotmarks}
\usetikzlibrary{arrows.meta}
\usepgfplotslibrary{patchplots}
\usepackage{amsmath}
\usepackage{subfig}
\usepackage{caption}

\renewcommand{\thetable}{\Roman{table}}
\def\tablename{TABLE}
\usepackage{booktabs,array}
\usepackage{multirow}
\usepackage{balance}
\usepackage{upgreek}

\renewcommand{\baselinestretch}{0.92} 

\def\BibTeX{{\rm B\kern-.05em{\sc i\kern-.025em b}\kern-.08em
    T\kern-.1667em\lower.7ex\hbox{E}\kern-.125emX}}

\definecolor{orangeSpec}{rgb}{1.000000,0.498039,0.054902}
\definecolor{greenSpec}{rgb}{0.329412,0.631373,0.000000}
\definecolor{blueSpec}{rgb}{0.121569,0.466667,0.705882}
\definecolor{redSpec}{rgb}{0.839216,0.152941,0.156863}
\definecolor{purpleSpec}{rgb}{0.49,0.18,0.56}
\definecolor{mycolor1}{rgb}{0.00000,0.50000,1.00000}%
\definecolor{mycolor2}{rgb}{0.00000,1.00000,1.00000}%
\definecolor{mycolor3}{rgb}{1.0000,0.80000,0.00000}%

\begin{document}

\title{A MIMO Radar-Based Metric Learning Approach for Activity Recognition}

	\author{\IEEEauthorblockN{Fady Aziz\textsuperscript{1}\IEEEauthorrefmark{1},
		Omar Metwally\textsuperscript{1}\IEEEauthorrefmark{1},
		Pascal Weller\textsuperscript{1},
		Urs Schneider\textsuperscript{1},
	Marco F. Huber\textsuperscript{2,3}}
	\IEEEauthorblockA{\textsuperscript{~1}Department of Bio-mechatronic Systems, Fraunhofer IPA, Stuttgart, Germany}
	\IEEEauthorblockA{\textsuperscript{~2}Institute for Industrial Manufacturing and Management, University of Stuttgart, Germany}
		\IEEEauthorblockA{\textsuperscript{~3}Center for Cyber Cognitive Intelligence, Fraunhofer IPA, Stuttgart, Germany}
				\thanks{Email: fady.aziz@ipa.fraunhofer.de, omarfekrymfm@gmail.com}
				\thanks{\IEEEauthorrefmark{1}\textbf{These authors equally contributed to this work.}}}
	
\maketitle
\begin{abstract}

\par Human activity recognition is seen of great importance in the medical and surveillance fields. Radar has shown great feasibility for this field based on the captured micro-Doppler (\(\upmu\)-D) signatures. In this paper, a MIMO radar is used to formulate a novel micro-motion spectrogram for the angular velocity (\(\upmu\)-\(\omega\)) in non-tangential scenarios. Combining both the \(\upmu\)-D and the \(\upmu\)-\(\omega\) signatures have shown better performance. Classification accuracy of \unit[88.9]{\%} was achieved based on a metric learning approach. The experimental setup was designed to capture micro-motion signatures on different aspect angles and line of sight (LOS). The utilized training dataset was of smaller size compared to the state-of-the-art techniques, where eight activities were captured. A few-shot learning approach is used to adapt the pre-trained model for fall detection. The final model has shown a classification accuracy of \unit[86.42]{\%} for ten activities.  
\end{abstract}

\begin{IEEEkeywords}
 Micro-angular velocity, activity recognition, MIMO radar, micro-Doppler, metric learning
\end{IEEEkeywords}

\section{Introduction}
\par Single-sensory solutions are heavily investigated nowadays for human activity recognition. For example, cameras and infrared have been investigated for different applications, e.g., gesture recognition \cite{song2018design ,sivamani2019design, 8698650}. Others have relied on wearable sensors to monitor human activity \cite{9278625, 8782506}. Among the presented studies, both visionary and wearable sensors have been investigated for fall detection, e.g., cameras \cite{waheed2017novel}, infrared \cite{9043000} and wearable sensors \cite{9278625}. Both categories of sensors can achieve acceptable performance, but other aspects, e.g., the vision-blocking conditions and the motion freedom, are challenging. Since radar sensors do not suffer from such limitations, many researchers were motivated to examine the radar feasibility for activity recognition.

\par Previous studies focused on using radar systems to capture the micro-motion signatures, which can uniquely reflect human activities. The most commonly used technique was capturing the micro motions on the Doppler dimension, known as the micro-Doppler (\(\upmu\)-D) signature \cite{chen2019micro}. For example, the walking \(\upmu\)-D signature is formulated as a time-frequency representation, where the micro swinging motions appear as superimposed frequency components to the main Doppler frequency component induced due to the translational motion \cite{chen2006micro}. Such \(\upmu\)-D signatures were utilized in many studies for human activity recognition and have shown acceptable classification accuracy \cite{zenaldin2016radar,jokanovic2016radar,zhang2017doppler,  klarenbeek2017multi,  erol2019gan,khasnobish2021novel,8283539}. As stated in \cite{gurbuz2019radar}, deep learning (DL) techniques show better performance than traditional learning techniques. However, it was stated that the size of the training dataset is one of the main challenges. In order to accommodate such a challenge, transfer learning was presented in \cite{gurbuz2019radar}, where simulated data were used for base training. 

\par Other solution was presented in \cite{erol2019gan}, where a generative adversarial network (GAN) was used for data augmentation. One of the main goals behind increasing the size of the training dataset was decreasing the classification confusion between similar activities, e.g., (bending and sitting) or (crawling and creeping). Another solution to solve the activities ambiguity was relying on more radar features, e.g., the range profiles \cite{erol2019radar}. The classification task in the aforementioned studies was based on a single occurrence of each activity. Thus, one solution is operating the radar to capture single snapshots separately, as presented in \cite{campbell2020attention,gurbuz2019radar} or slicing a long-period captured signature \cite{li2019activities}. However, the activities are expected to occur in sequence in realistic scenarios, which will require adaptive slicing, e.g., the recurrent neural network (RNN) proposed in \cite{gurbuz2019radar}. Also in \cite{li2019activities}, streams of multiple activities were sliced based on a sliding window with an overlap of \unit[50-70]{\%}. There are also other aspects to be considered such as the data generalization and the detection aspect angle \cite{gurbuz2019radar}.

\begin{figure*}
	\centering
	\centerline{\includegraphics[width=\textwidth]{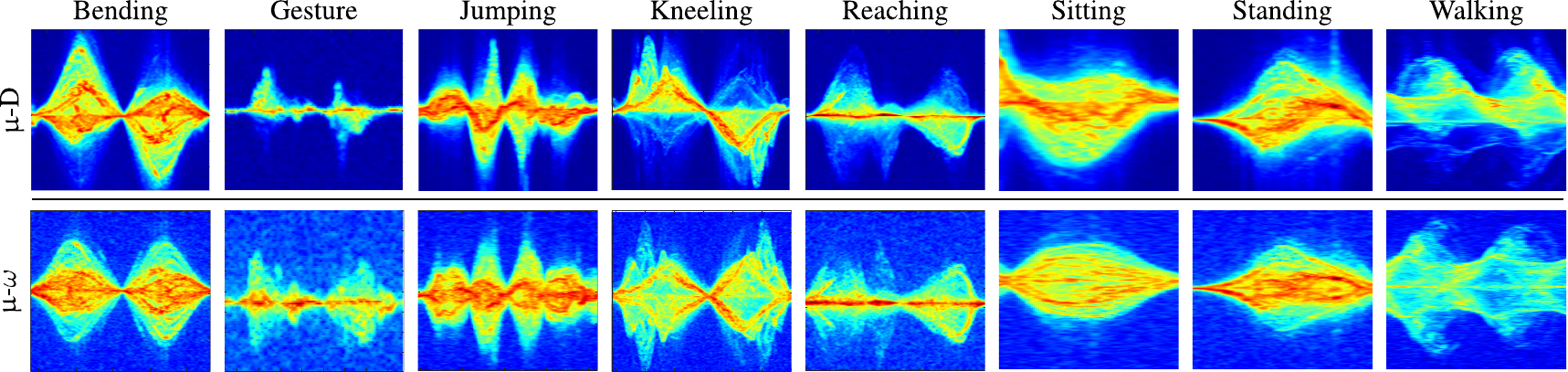}}
	\caption{\centering The \(\upmu\)-D and \(\upmu\)-\(\omega\) signatures for the main $8$ activities.}
	\vspace{-4mm}
\end{figure*}

\par  The body micro motions can also be described as the angular displacements of a nonrigid body in space \cite{chen2006micro,chen2019micro}. Therefore, relying on a multiple-input-multiple-output (MIMO) radar can be utilized for capturing the angle of arrival (AoA) trajectories of different body parts that can be utilized for activity realization, such as the gesture recognition presented in \cite{skaria2019hand}. The AoA is seen of great feasibility due to the growing development in MIMO modules regarding availability, size, and detection resolution. Moreover, it can be processed to reflect the micro angular velocity (\(\upmu\)-\(\omega\)). To the best of our knowledge, this is not utilized except in our previous study, presented in \cite{weller2021mimo} for human identification. This is different from the technique presented in \cite{nanzer2010millimeter}, where the tangential angular velocity is estimated based on interferometric frequency with no AoA estimation. Such a technique is not feasible for non-tangential scenarios, as the estimated velocity component is complementary to the radial velocity \cite{nanzer2010millimeter}. 

\par In this paper, a novel technique for estimating the (\(\upmu\)-\(\omega\)) signature is presented, based on calculating the rate of change of the AoA with respect to time. The derived \(\upmu\)-\(\omega\) signature reflects the micro-motion behavior and is time-synchronized with the \(\upmu\)-D signature. Relying on both micro-motion signatures is investigated to decrease the classification confusion between similar activities. The study is conducted on eight different activities, and the experimental setup is designed to test different aspects, which are the data generalization, the detection aspect angle, and the training data size. The classification task is based on a metric learning approach using a triplet loss function. A FSL approach is implemented to include \textit{falling} and \textit{standing from falling}, which are considered as hard-to-collect activities. A frequency modulation continuous wave (FMCW) radar with MIMO configuration operating at \unit[77]{GHz} is utilized and parameterized with the maximum achievable angular resolution to detect activities with small cross-section, e.g., gesture and reaching.       

\vspace{-0.3mm}
\section{Micro-motion Signatures Formulation}
\par The human body is considered as a nonrigid body, whose motion causes a change in the body shape. Any non-translational micro-motion, e.g., limbs motion while walking, is realized by the radar as micro velocity components \cite{chen2019micro}. Such radial micro velocities are analyzed through time-frequency representation to formulate the \(\upmu\)-D signatures. However, the human micro motions induce angular displacement, in which \(\upmu\)-\(\omega\) signatures can be formulated in a similar way to reflect the micro-motion behavior. Therefore, similar to analyzing the range profiles through time for deriving the \(\upmu\)-D signatures, the angle profiles can be processed similarly. Realizing the angular velocity in such way is more feasible than the technique presented in \cite{nanzer2010millimeter}, as it does not have any hardware obligation. Unlikely, the technique presented in \cite{nanzer2010millimeter} requires widely-spaced receiving antennas. Moreover, processing on the captured range-AoA maps by the MIMO radar, enables simultaneous formulation of both micro-motion signatures. Unlikely relying on the other technique will offer only one signature at a time, due to the complementary relation between both velocity components\cite{nanzer2010millimeter}. Accordingly, the technique presented in \cite{nanzer2010millimeter} solves the problem of decayed \(\upmu\)-D signature in tangential scenarios \cite{wang2019simultaneous,nanzer2016micro,liang2020enhanced}, without adding extra feature to the radial \(\upmu\)-D signature. 

\par The MIMO configuration is based on including spatially-spaced multiple receivers with a distance ($d_R = \lambda/2$), where $\lambda$ is the transmission wavelength. The FMCW transmission protocol is used for range estimation, while the interferometric analysis between different receivers is used for AoA estimation \cite{geibig2016compact}. As stated in \cite{milligan2005modern}, the \unit[3]{dB} angular resolution and the estimated AoA are described as follows: 
\begin{equation}
\theta_{res} = \frac{1.78}{N_{Rx}}, \hspace{1.2 cm} \theta = arcsin\frac{\lambda\Delta\epsilon}{2\pi d_R}
\end{equation}

\par where $N_{Rx}$ is the number of receivers, and $\Delta\epsilon$ is the phase difference between adjacent receivers. Accordingly, a Range-AoA map representing sagittal plane scanning is formulated within each chirp transmission period ($T_c$) \cite{geibig2016compact}, and is accumulated through time. Then, a short-time Fourier transform (STFT) is applied on both range and AoA dimensions. Such operation yields time-frequency spectrograms for both the \(\upmu\)-D and the \(\upmu\)-\(\omega\) signatures, respectively. To analyze the newly-formulated \(\upmu\)-\(\omega\) signature, the walking activity was selected as all the body parts contribute with periodic behavior \cite{chen2019micro}. A study about the feasibility of utilizing that signature for walking human identification is presented by us in \cite{weller2021mimo}. 

\begin{figure}
	\centering
	\hspace{-4mm}\resizebox{0.95\columnwidth}{!}{
		\input{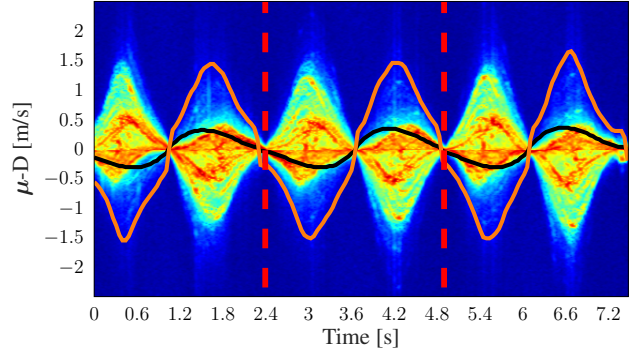}
	}
	
	\caption{\centering An example for the adaptive extraction of single instant of bending activity.}
	\label{fig:slicing}
	\vspace{-6mm}
\end{figure}

\section{Dataset Preparation} \label{sec:dataset}
\subsection{Experimental Setup}

\label{sec:expSetup}
\par In this paper, ten activities are selected for recognition that are crucial for home applications and heavily investigated in other studies \cite{erol2019gan,gurbuz2019radar}. Fig.~\ref{fig:slicing} shows both the \(\upmu\)-D and the \(\upmu\)-\(\omega\) signatures for the different activities, which can be categorized as follows: 
\begin{itemize}
	\item Sitting, standing, and walking, which usually occur in sequence and include contribution from the whole body.  
	\item Gesture and reaching, which are vital for smart home applications and include contribution from the arm only. 
	\item Bending, kneeling, and jumping, which include contribution from the lower body. 
	\item Falling and standing from falling, as fall detection is vital for the elderly.  
\end{itemize}

\begin{figure}
	\captionsetup[subfigure]{oneside,margin={1.cm,0cm}}
	\subfloat[\(\upmu\)-D signature.\label{fig:fd}]{\resizebox{0.98\columnwidth}{!}{
%
%

\begin{tikzpicture}
\begin{axis}[%
width=4.521in,
height=1.6in,
at={(0in,0in)},
scale only axis,
point meta min=-90.8275160996101,
point meta max=-28.6200534312793,
axis on top,
xmin=0.0,
xmax=4.5,
xtick distance=0.4,
xlabel style={font=\color{white!15!black},yshift=0.5mm},
xlabel={\Large Time [s]},
ymin=-3.4,
ymax=3.4,
ytick distance=1,
ytick style={draw=none},
xtick style={draw=none},
yticklabel style = {font=\large,xshift=-1mm},
xticklabel style = {font=\large,yshift=-1mm},
ylabel style={font=\color{white!15!black}, yshift=0mm},
ylabel={\Large \(\upmu\)-D [m/s]},
axis background/.style={fill=white},
colormap/jet
]
\addplot [forget plot] graphics [xmin=-0.00542667076700434, xmax=30.004062670767, ymin=-4.60030396979182, ymax=4.58236926798677] {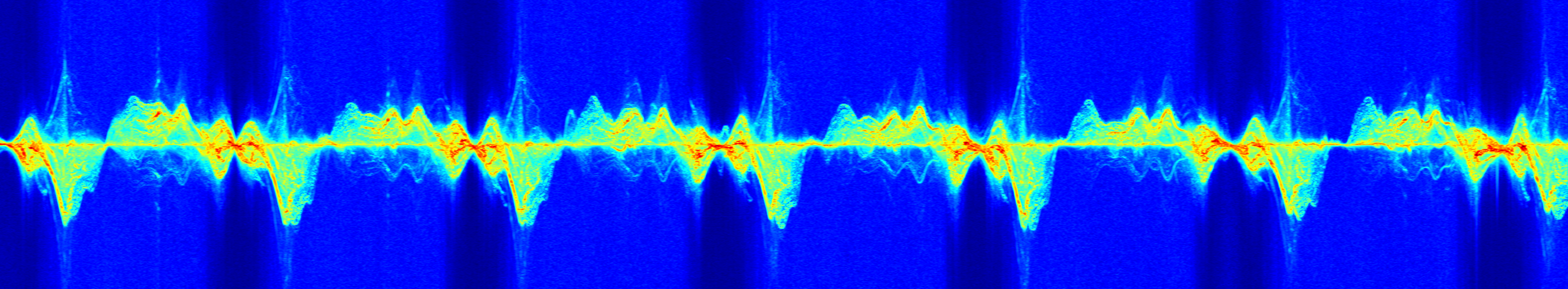};


\draw [white,ultra thick,decoration={brace,amplitude=15pt,raise=5pt},decorate]
(axis cs:0.2,1.0) --
node[above=20pt] {\color{white} \Large \textbf{Falling}} 
(axis cs:1.9,1.0);

\draw [white,ultra thick,decoration={brace,amplitude=15pt,raise=5pt,mirror},decorate]
(axis cs:2.2,-1.0) --
node[align=center,midway,anchor=north,yshift=-5ex] {\color{white} \Large \textbf{Standing after falling}} 
(axis cs:4.4,-1.0);

\draw [red, line width=1.25mm, dashed, dash pattern=on 12pt off 12pt](axis cs:2.025,-3.4) -- (axis cs:2.025,3.4);
\end{axis}
\end{tikzpicture}%
}}
	\vspace{2mm}
	\subfloat[\(\upmu\)-\(\omega\) signature.\label{fig:fa}]{\resizebox{0.98\columnwidth}{!}{
%
%
\begin{tikzpicture}
\begin{axis}[%
width=4.521in,
height=1.6in,
at={(0in,0in)},
scale only axis,
point meta min=-90.8275160996101,
point meta max=-28.6200534312793,
axis on top,
xmin=-0.00542667076700434,
xmax=4.49388932250372,
xtick distance=0.4,
ytick distance=0.4,
xlabel style={font=\color{white!15!black},yshift=0.5mm},
xlabel={\Large Time [s]},
ymin=-1.05,
ymax=1.05981110128123,
ytick style={draw=none},
xtick style={draw=none},
yticklabel style = {font=\large,xshift=-1mm},
xticklabel style = {font=\large,yshift=-1mm},
ylabel style={font=\color{white!15!black}, yshift=-2mm},
ylabel={\Large \(\upmu\)-\(\omega\) [rad/s]},
axis background/.style={fill=white},
colormap/jet,
]
\addplot [forget plot] graphics [xmin=-0.00542667076700434, xmax=30.004062670767, ymin=-1.32298607488323, ymax=1.32298607488323] {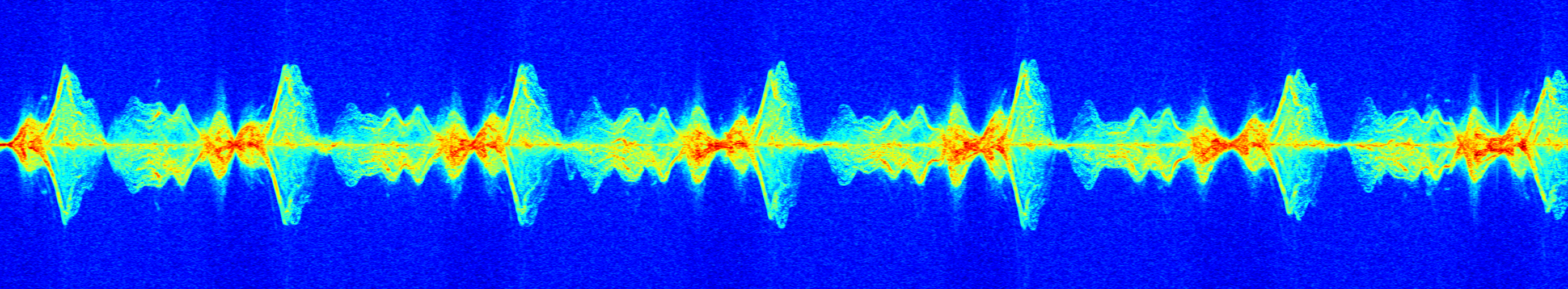};

\draw [white,ultra thick,decoration={brace,amplitude=15pt,raise=5pt},decorate]
(axis cs:0.2,0.32) --
node[above=20pt] {\color{white} \Large \textbf{Falling}} 
(axis cs:1.9,0.32);

\draw [white,ultra thick,decoration={brace,amplitude=15pt,raise=5pt,mirror},decorate]
(axis cs:2.2,-0.3) --
node[align=center,midway,anchor=north,yshift=-5ex] {\color{white} \Large \textbf{Standing after falling}} 
(axis cs:4.4,-0.3);

\draw [red, line width=1.25mm, dashed, dash pattern=on 12pt off 12pt](axis cs:2.025,-2.99763637283667) -- (axis cs:2.025,3.5);
\end{axis}
\end{tikzpicture}%
}}
	\caption{\centering Falling/Standing from falling signatures. \label{fig:fsf}}
	\vspace{-4mm}
\end{figure}
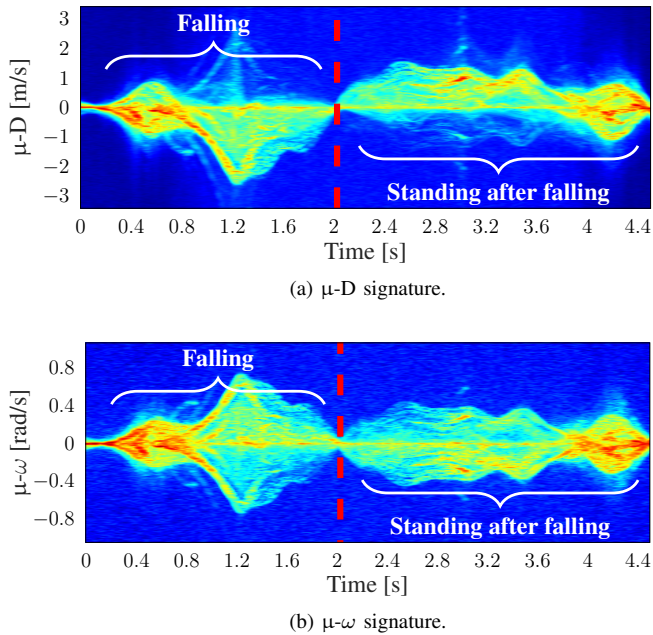

\par The signatures were captured on 8 different human subjects with height range of \unit[155-192]{cm} and weight range of \unit[70-105]{Kg} to assure generalization. Each subject has performed each activity in front of the radar for \unit[30]{s} at a distance range of \unit[1-3]{m}, except the walking activity, which was held at a distance range of \unit[1-5]{m} from the radar. Due to the fact that detection at aspect angles curtails the radar signals \cite{tahmoush2009angle}, the data was collected at different aspect angles, which are line of sight (LOS), $\pm\unit[30]{^\circ}$ and $\pm\unit[50]{^\circ}$ to assure pervasive recognition. To ensure data generalization and avoid redundancy, the detection range and the aspect angles were not fixed for all targets. Thus, each subject was asked to perform each activity at four locations for training and another two for test at the line of sight and randomly selected aspect angles of $\pm\unit[30]{^\circ}$ or $\pm\unit[50]{^\circ}$. Both the \(\upmu\)-D and the \(\upmu\)-\(\omega\) signatures were recorded, then different techniques were applied to slice the spectrograms into single occurrences of each activity. The sitting and standing activity were collected in sequence, and the recording time was doubled to ensure balanced data for all classes. The falling and standing from falling were also collected in sequence and resulted in fewer data samples compared to other activities, due to the experimental complexity. The last two activities weren't collected at different aspect angles, but instead each subject used to fall and stand in sequence at the four sides of the body (front, back, left and right sides). The same period of \unit[30]{s} is used for each record. 

\par Since including more $N_{Rx}$ yield a better angular resolution, both transmitting antennas were used to simulate the effect of $2 \times N_{Rx}$ \cite{pirkani2019implementation}, resulting in an angular resolution of $\theta_{res}=\unit[3.19]{^\circ}$. Thus, activities as gesture and reaching, which require high angle and range resolutions, can be detected. The radar is parameterized as shown in Table~\ref{tab:param}. The last two activities, which are falling and standing from falling are collected to test the feasibility of the metric learning approach for FSL. The same period of \unit[30]{s} was recorded for each subject, revealing much fewer data samples as the full sequence of falling and standing lasts for $\approx \unit[5]{s}$ as shown in Fig.~\ref{fig:fsf}, unlike single occurrence of other activities that occurs in $\approx \unit[1.5]{s}$. Both the falling and standing from falling were collected in sequence within each capturing period of \unit[30]{s}. 

\begin{table}[!b]
	\vspace{-3mm}	
	\setlength\arrayrulewidth{0.01pt}
	\def\arraystretch{0.9}
	\scriptsize
	\centering
	\caption{\centering MIMO radar module parametrization.  \label{tab:param}\vspace{-2mm}}
	\resizebox{\columnwidth}{!}{
		\begin{tabular}{l r l r}
			\toprule
			\multicolumn{2}{c}{Radar Parametrization} & \multicolumn{2}{c}{Attributes} \\
			\midrule
			Carrier frequency ($f_o$)  & \unit[77]{GHz} & {} & {}\\
			Tx-Rx antennas  & 2-16 & $\theta_{res}$ & \unit[3.19]{$^\circ$}\\
			Bandwidth ($B$)  & \unit[0.25]{GHz} & {} & {}\\
			Chirp duration ($T_{c}$)   & \unit[80]{$\mu s$} & $v_{res}$ & \unit[4.753]{cm/s}\\
			Samples per chirp ($N_S$)   & 112 & {} & {}\\
			Chirps per frame ($N_P$)   & 512 & {} & {}\\
			\bottomrule
		\end{tabular}
	}
\end{table}

\subsection{Slicing Techniques}\label{Sec:slice}
\begin{figure}[t!]
	\centering
	\centerline{\includegraphics[width=0.475\textwidth]{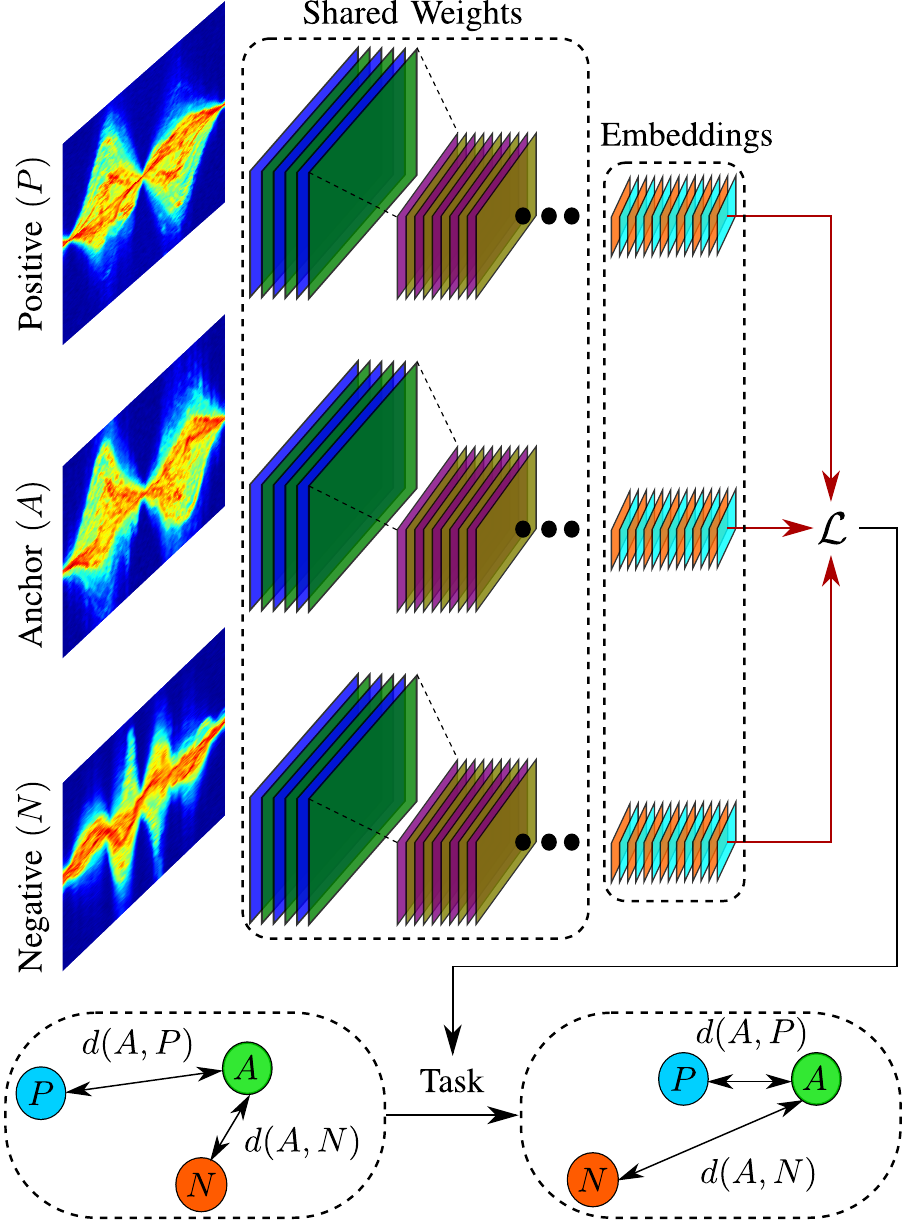}}
	\caption{\centering Metric learning model architecture.}
	\label{fig:2}
	\vspace{-6mm}
\end{figure}
\begin{figure*}[t!]
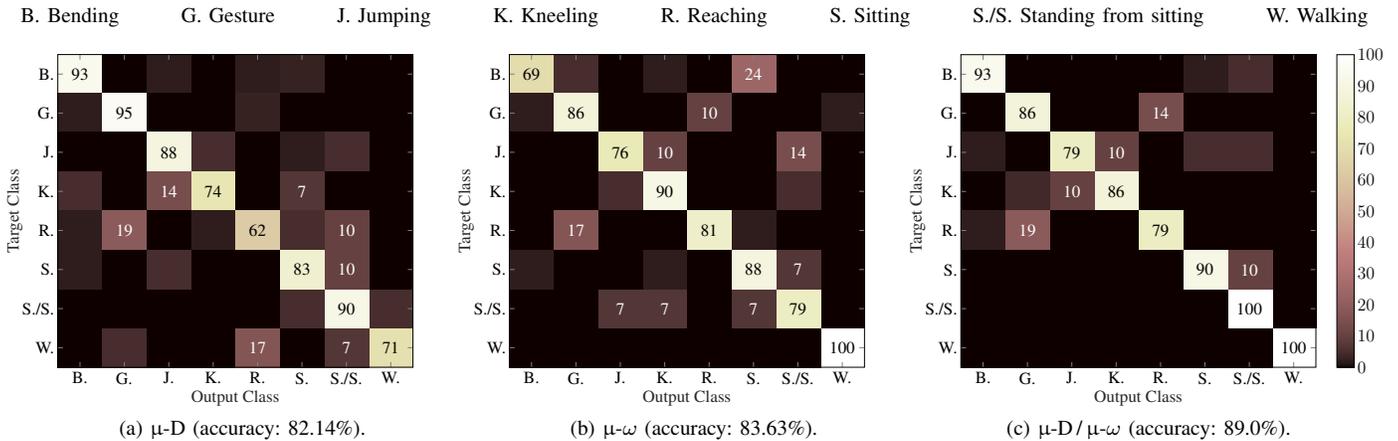

	\hspace{1mm} \footnotesize B. Bending\hfill G. Gesture\hfill J. Jumping\hfill K. Kneeling\hfill R. Reaching\hfill S. Sitting\hfill S./S. Standing from sitting\hfill W. Walking
	\vfill
	\vspace{-1mm}
	\captionsetup[subfigure]{oneside,margin={1.cm,0cm}}
	\subfloat[\(\upmu\)-D (accuracy: 82.14\%). \label{fig:conf_doppler}]{\resizebox{0.3\textwidth}{!}{\input{Figures/conf_Dopp.tex}}}
	\qquad
	\captionsetup[subfigure]{oneside,margin={1.cm,0cm}}
	\subfloat[\(\upmu\)-\(\omega\) (accuracy: 83.63\%). \label{fig:conf_angle}]{\resizebox{0.3\textwidth}{!}{\input{Figures/conf_Angle.tex}}}
	\qquad
	\captionsetup[subfigure]{oneside,margin={0.15cm,0cm}}
	\subfloat[\(\upmu\)-D$\,$/$\,$\(\upmu\)-\(\omega\) (accuracy: 89.0\%). \label{fig:conf_da}]{\resizebox{0.3515\textwidth}{!}{\input{Figures/conf_DA.tex}}}
	\caption{\centering Confusion matrix for all the micro-motion signatures constellations. \label{fig:deviations_all}}
	\vspace{-6mm}
\end{figure*}
\par For extracting a single occurrence of each activity, three techniques were applied to the captured spectrograms. First, a fixed-time window of \unit[1.5]{s} is used, as that was observed to be the average duration for a single cycle for most of the activities. Such technique is not with high feasibility for realistic scenarios, as the different activities are expected to occur in sequence, and the single slice may include half cycles from two different activities or silent period.  For better data analysis, a sliding window of \unit[1.5]{s} was applied. Such a technique acts as a data augmentation technique, where the network is guaranteed to see the full cycle of each activity from different views, and the training data size is increased. The third technique is adaptive slicing that extracts a single occurrence of each activity accurately, as shown in Fig.~\ref{fig:slicing}. It is based on detecting the center of gravity (CoG) behavior in the captured spectrogram based on the algorithm presented in \cite{9093181}. The CoG is used to determine the direction of movement, which is combined with the motion cycles envelope to determine the slicing locations. Moreover, the slicing frequency is tunable and is fixed for each activity to ensure generalization. 

\section{Metric Learning}
\par Metric learning models are deep learning architecture consisting of two or more parallel and identical networks. The main goal is to learn a similarity that maps the input feature map to a latent space \cite{zhang2016siamese}. This architecture does not require to be trained with huge datasets, which makes it feasible for radar-based activity recognition, e.g., human identification \cite{ni2021open,yang2019person}. Thus, metric learning is of great feasibility for applications as fall detection, where a small training dataset is available, and FSL is required. For our presented study, the model is first trained on the primary eight activities and is able to adapt to two unseen activities without retraining. The utilized model is composed of a 10-layer identical triplet network, taking as an input 256$\times$128 resized gray-scale images. The network is composed of 6 convolution layers followed by flatten layer and 4 dense layers with 0.3 dropout. It ends with L2 normalized embeddings for triplet loss calculation. 

\subsection{Triplet Loss}
\par The classification task is based on the triplet loss function. It is based on taking 3 samples as input, which are positive ($P$), anchor ($A$) and negative ($N$). Both the $P$ and the $A$ belong to the same class, while the $N$ belongs to other class. The triplet loss can be described as follows: 
\begin{equation}
\ell(A,P,N) = \max[d(A,P)-d(A,N)+\delta,0]
\vspace{-1mm}
\end{equation}
\par where $d$ is the Euclidean distance and \(\delta\) represents a tunable margin. The main goal is minimizing the loss by achieving \(d(A,P)\rightarrow0\) and \(d(A,N)\rightarrow\:>d(A,P)+\delta\). Mining the batch-hard triplets occurs online within the training process to assure educational training. The mini-batches are settled to be balanced within the same process to assure an unbiased training result. Thus, the experimental setup is conducted on different aspect angles with respect to the radar to test the ability of the transfer learning for deriving a correlation between the captured signatures on different angles. 

\begin{figure*}
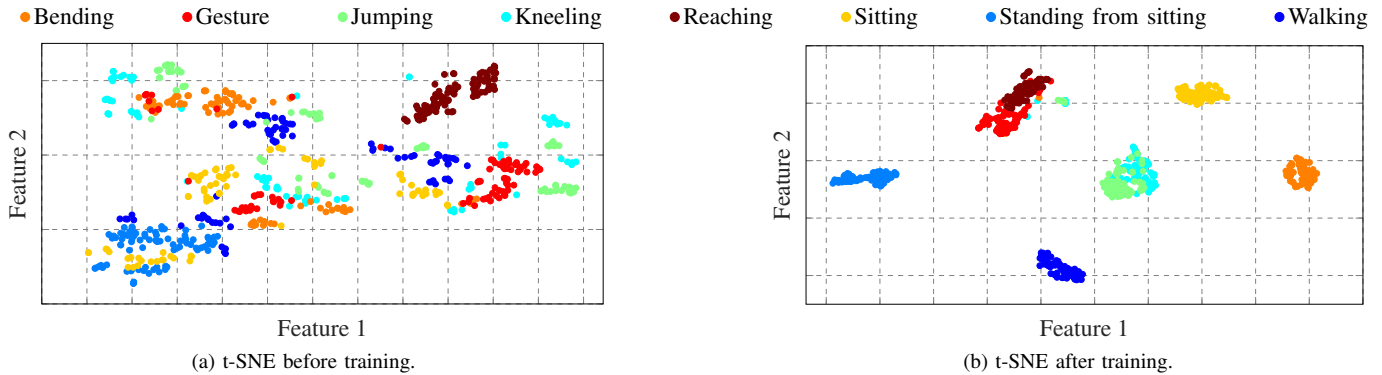

	\hspace{1mm} \small \ref{p:s}$\,$Bending\hfill \ref{p:ss}$\,$Gesture\hfill \ref{p:k}$\,$Jumping\hfill \ref{p:j}$\,$Kneeling\hfill \ref{p:w}$\,$Reaching\hfill \ref{p:r}$\,$Sitting\hfill \ref{p:g}$\,$Standing from sitting\hfill \ref{p:b}$\,$Walking
	\vfill
	\vspace{-2mm}
	\centering
	\centerline{\subfloat[t-SNE before training.]{\resizebox{.9\columnwidth}{!}{\input{Figures/tsne_before.tex}}\label{fig:t1}}
		\hfill
		\subfloat[t-SNE after training.]{\resizebox{.9\columnwidth}{!}{\input{Figures/tsne_after.tex}}\label{fig:t2}}}
	\caption{\centering t-SNE of the embedding space for the adaptive-sliced and combined \(\upmu\)-D and \(\upmu\)-\(\omega\) signatures. \label{fig:tsne}}
	\vspace{-4mm}
\end{figure*}
\section{Results and Discussion}
\subsection{Slicing Techniques}
\par The performance of the three slicing techniques is compared in terms of the training data size, the data consistency, and the classification accuracy. The proposed network is trained on the resulting \(\upmu\)-D sliced signatures, as it is the most commonly used. All the reported classification accuracy is on the test dataset for all the activities. The test dataset was collected separately, as described in Sec.~\ref{sec:expSetup}. As shown in Table~\ref{tab:dataset1}, the first technique of applying constant window of \unit[1.5]{s} yields the minimum number of samples. Although the \unit[1.5]{s} is estimated to be the average duration of a single occurrence for all the activities, such technique does not show a high level of consistency for the sliced samples, especially for the sitting and standing that were collected in sequence. Thus, such a technique yields the least classification accuracy of \unit[53.43]{\%}. 

\par The second technique of applying a sliding window of \unit[1.5]{s} with an overlap of \unit[80]{\%} yields a total of 6288 augmented dataset. This technique has shown an overall accuracy of \unit[78.22]{\%}. This is nearly the same accuracy achieved for the augmented dataset presented in \cite{erol2019gan} for similar activities and on nearly the same training dataset size. Moreover, our metric learning approach has shown comparable performance to the study presented in \cite{li2019activities}, which is based on using SVM with a similar sliding window algorithm. However, such a study has a real-time limitation as the sliding window is of \unit[3-5]{s} with an overlap of \unit[70]{\%}. Both our study and the presented study in \cite{li2019activities} are showing more confusion for sitting with standing, which are captured in sequence as they are supposed to occur consecutively. Thus, adaptive slicing is required to decrease that classification ambiguity.  

\par The third technique introduces the idea of adaptive slicing for each activity, which simulates the behavioral slicing. This technique has shown the best classification accuracy of \unit[82.14]{\%} on a training dataset of 680 samples, which are of comparable dataset size to the utilized dataset for the transfer learning approach, presented in \cite{gurbuz2019radar}. Although our study isn't conducted on all the 12 activities presented in \cite{gurbuz2019radar}, our proposed metric learning approach is still giving a good indication due to the limited available dataset. Moreover, the activities were captured on different aspect angles, unlike most of the previous studies that included only LOS. Additionally, the adaptive slicing technique shows great feasibility in slicing the streams of standing and sitting, which can be extended to streams of multiple activities. 

\begin{table}[b!]
	\vspace{-3mm}
	\setlength{\tabcolsep}{ 5 pt}
	\scriptsize
	\centering
	\caption{\centering Impact of slicing techniques on the classification performance. The results are based on the test dataset.}
	\resizebox{\columnwidth}{!}{
		\begin{tabular}{lrrr}
			\toprule
			\multirow{2}{*}{} & \multicolumn{3}{c}{Slicing technique}       \\
			\cmidrule{2-4}
			& Time (discrete) & Time (sliding) & Adaptive \\
			\midrule
			Samples/Class     & 60              & 786            & 85       \\
			Total data        & 472             & 6288           & 680      \\
			Accuracy [\%]          & 53.43           & 78.22          & 82.14   \\
			\bottomrule
		\end{tabular}
	}
	\label{tab:dataset1}
\end{table}

\subsection{Micro-motion Signatures Constellations}
\par Since the adaptive slicing has shown the best classification performance for the \(\upmu\)-D signatures, it is used for slicing the \(\upmu\)-\(\omega\) as well, since both signatures are time-synchronized. Training on the \(\upmu\)-\(\omega\) only resulted in an overall accuracy of \unit[83.63]{\%}. As shown in Fig.~\ref{fig:deviations_all}, relying only on either of the \(\upmu\)-D or the \(\upmu\)-\(\omega\) will cause a classification drop on some activities. For example, in case of the \(\upmu\)-D, jumping is confused with kneeling, and reaching is confused with gesture and standing from sitting. While for the \(\upmu\)-\(\omega\), bending is confused with sitting, and jumping is confused with both kneeling and standing from sitting. Accordingly, another training was held based on combining both the \(\upmu\)-D and the \(\upmu\)-\(\omega\) together. As shown in Fig.~\ref{fig:conf_da}, such training has shown the best performance of \unit[89]{\%}, with classification accuracy for each class of $\geq\unit[80]{\%}$.     

\par The quality of the embedding space affects the performance of the metric learning. A t-distributed stochastic neighbor embedding (t-SNE) algorithm can be used for the embedding space visualization, which is shown in Fig.~\ref{fig:tsne}. The complexity of the dataset can be observed as all the classes are confused with each other, as shown in Fig.~\ref{fig:t1}. On the other hand, our proposed network resulted in well-separated classes, as shown in Fig.~\ref{fig:t2}. The t-SNE representation agrees with the confusion matrix of the combined micro-motion signatures, shown in Fig.~\ref{fig:conf_da}. The bending, walking, standing, and sitting form separate clusters, while each of the gesture forms a cluster with reaching and similarly jumping with kneeling. Thus, it can be observed that the metric learning approach could decrease the inter-class variances.

\subsection{Few-Shot Learning} 
\begin{figure}[t!]
	\centering
	\resizebox{0.95\columnwidth}{!}{
		\input{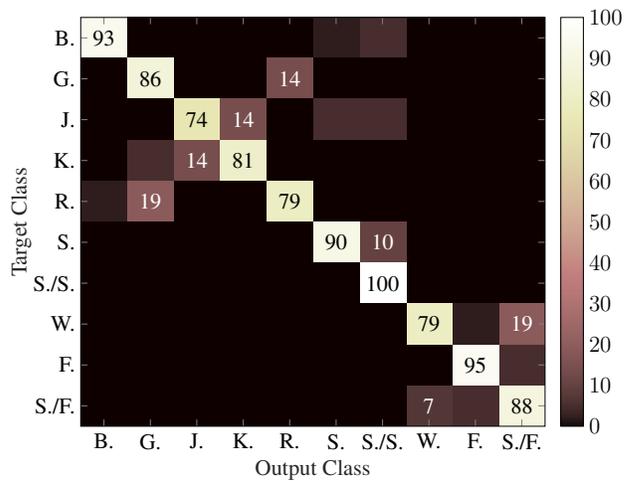}
	}
	\caption{\centering Confusion matrix for the few-shot learning  including the falling (F.) and standing from falling (S./F.)}
	\label{fig:conffew}
	\vspace{-6mm}
\end{figure}
\par FSL has shown acceptable performance for (\(\upmu\)-D)-based human identification based on walking activity \cite{ni2021open,yang2019person}. The human identification task requires multitudes of data as the classification is required to be done on the same activity for many subjects, where the FSL and transfer learning were presented as remedies for collecting huge datasets. Similarly, differentiating between multiple activities suffers from the same problems. In our study, falling and standing from falling have been selected to test the feasibility of FSL. Fall detection is considered of high importance for the elderly, where radar has been foreseen of great feasibility as it is preserving the privacy aspect and can be mounted anywhere, e.g., in bathrooms. Other studies have included falling as a primary activity, which required collecting more data compared to our technique. The adaptive slicing technique has shown great consistency in slicing both activities in sequence. 

\par The network trained on the eight activities is fine-tuned to include the new two classes. The training was done on only 15 samples for each of the two classes based on combining both the \(\upmu\)-D and \(\upmu\)-\(\omega\) signatures. The test dataset was balanced between the ten classes, in which each class includes 42 samples. The FSL procedure resulted in a classification accuracy of \unit[86.42]{\%}. As shown in Fig.~\ref{fig:conffew}, both falling and standing from falling can be classified with an accuracy of \unit[95]{\%} and \unit[88]{\%}, respectively. Moreover, the classification accuracy of the eight primary classes was not affected when the last two activities were included.

\section{Conclusion and Future Work}
\par In this paper, a metric learning approach based on the triplet loss function is presented for radar-based human activity recognition. A MIMO radar is utilized, where the performance of combining a newly-formulated \(\upmu\)-\(\omega\) signature with the commonly-used \(\upmu\)-D signature is tested. The study is conducted on eight vital activities for home applications. The experimental setup is based on capturing the data on multiple aspect angles in a range of [$\pm\unit[30]{^\circ}$-$\pm\unit[50]{^\circ}$] and LOS. The data was collected in periods of \unit[30]{s} for different subjects, and three slicing techniques were tested, which are a constant window of \unit[1.5]{s}, a sliding window with an overlap of \unit[80]{\%} and adaptive slicing based on the behavior of each activity. Both the \(\upmu\)-D and the \(\upmu\)-\(\omega\) resulted in comparable classification performance of \unit[82.14]{\%} for the \(\upmu\)-D and \unit[83.63]{\%} for the \(\upmu\)-\(\omega\). Combining both micro-motion signatures shows the best accuracy of \unit[89]{\%}. Finally, the FSL is trained on falling and standing from falling with only 15 samples for each activity and resulted in a classification accuracy of \unit[86.42]{\%} for the ten activities. 

\par In this study, no activity sequence is proposed while capturing the data as it was shown in \cite{li2019activities} that it doesn't complicate the classification task as long as the activities are not similar. However, the robustness of the adaptive slicing will be tested on a sequence of activities as future work. Moreover, techniques as RNN proposed in \cite{gurbuz2019radar} should be applied to both micro-motion signatures on a real-time basis. Since our study is based on a MIMO radar module, scenarios that include multiple targets can be included. The idea of FSL should be investigated for more activities and extended to open set recognition.  

\vspace{12pt}

\bibliographystyle{IEEEtran}

\end{document}